\documentclass[a4paper,11pt,hyper]{JHEP3}

\usepackage{graphicx}% Include figure files
\usepackage{dcolumn}% Align table columns on decimal point
\usepackage{bm}% bold math
\usepackage{latexsym}
\usepackage{amsfonts}
\usepackage{amssymb}
\usepackage{amsmath}
\def\ifempty#1{\def\tmpdata{#1}\ifx\tmpdata\empty }

\def\linebreak{\hfill\break}

\def\bra<#1|{\langle #1\rvert}
\def\ket|#1>{\lvert#1 \rangle}
\def\braket<#1|#2>{\langle #1|#2 \rangle}

\def\otop#1{\hbox{$#1\kern-0.1em$\llap{\hbox{\raise1.7ex\hbox{$\scriptstyle\circ$}}}} }

\def\inpare#1{\left(#1\right)}
\def\bigpare(#1){\left(#1\right)}
\def\inrbra#1{\left\{ #1 \right\}}

\def\inang#1{\langle {#1} \rangle}
\def\EXP#1{\inang{#1}}
\def\bigbra[#1]{\left[ #1 \right]}

\def\b{\bar }

\def\tend{\rightarrow}

\def\therefore{\mbox{\setbox0=\hbox{X}\hbox{$\ldotp$}\raise0.7\ht0\hbox{$\ldotp$}\hbox{$\ldotp$}} \quad }
\def\because{\mbox{\setbox0=\hbox{X}\raise0.7\ht0\hbox{$\ldotp$}\hbox{$\ldotp$}\raise0.7\ht0\hbox{$\ldotp$}}\kern0pt }

\def\bm#1{\boldsymbol{#1}}

\def\upin{\hbox{\setbox0=\hbox{$\cup$} \vrule width 0.05 \wd0 height \ht0 depth 0pt \kern - 0.5\wd0 \box0 }}
\def\Frac(#1/#2){\left(\frac{#1}{#2}\right)}

\def\sdprod{\mathrel{{\setbox0=\hbox{$\displaystyle\times$}\lower0.3\wd0\hbox{$\stackrel{\box0}{\scriptstyle\sim}$}}}}

\def\tosigma#1,{%
    \ifx\tmpindex\relax \def\tmpindex{#1} \let\next=\tosigma
    \else \ifnum\tmpindex=0 1 \else \sigma_\tmpindex \fi
          \ifx#1\relax  \let\next=\relax
          \else \otimes \let\next=\tosigma \def\tmpindex{#1} \fi
    \fi \next}
\def\tspb(#1){\let\tmpindex=\relax\tosigma#1,\relax,}
\def\HyperG(#1,#2;#3;#4){F\inpare{\textstyle #1,#2;#3;#4}}
\def\Eq#1{\begin{equation} #1 \end{equation}}

\def\Eqr#1{\begin{eqnarray} #1 \end{eqnarray}}

\def\Eqrsub#1{\begin{subequations}\Eqr{#1}\end{subequations}}
\def\Eqrsubl#1#2{\begin{subequations}
  \expandafter\ifx\csname Rlabel\endcsname \relax \label{#1}
  \else \Rlabel{#1} \fi \Eqr{#2}\end{subequations}}
\def\Bitm{\begin{itemize}}
\def\Eitm{\end{itemize}}
\def\Blist#1#2{\begin{list}{#1}{\parsep=0pt \itemsep=0pt%   
  \listparindent=0pt #2}}
\def\Elist{\end{list}}
\long\def\ignore#1#2{\def\ignoreflag{#1}\long\def\tmptext{#2}
  \ifnum\ignoreflag>1 #2 \fi}

\title{Parity Violation in Graviton Non-gaussianity}
\author{
Jiro Soda\\
Department of Physics, Kyoto University, Kyoto, 606-8502, Japan\\
 \email{jiro@tap.scphys.kyoto-u.ac.jp} 
}
\author{
Hideo Kodama\\
Theory Center, KEK, Tsukuba 305-0801, Japan;\\
Department of Particles and Nuclear Physics, 
The Graduate University for Advanced Studies, Tsukuba 305-0801, Japan\\ 
\email{Hideo.Kodama@kek.jp} 
}
\author{
Masato Nozawa\\
Theory Center, KEK, Tsukuba 305-0801, Japan\\
\email{nozawam@post.kek.jp}
}

\abstract{
We study parity violation in graviton non-gaussianity generated during inflation. We develop a useful formalism to calculate graviton non-gaussianity. Using this formalism, we explicitly calculate the parity violating part of the bispectrum for primordial gravitational waves in the exact de Sitter spacetime and prove that no parity violation appears in the non-gaussianity. We also extend the analysis to slow-roll inflation and find that the parity violation of the bispectrum is proportional to the slow-roll parameter. We argue that parity violating non-gaussianity can be tested by the CMB. Our results are also useful for calculating three-point function of the stress tensor in the non-conformal field theory through the gravity/field theory correspondence.}

\preprint{KUNS-2346, KEK-TH-1467, KEK-Cosmo-76}

\maketitle

\begin{document}

%-----------------------------------------------%
%                                               %
%               Introduction                    %
%                                               %
%-----------------------------------------------%

%T1>Introduction
\section{Introduction}

The origin of the chirality of the matter sector is one of the most
basic problems in particle physics.  The associated violation of
parity (P) and charge conjugation (C) invariance, however,  may not be so
fundamental because such a violation can be easily realized by
introducing a P violating hidden background field or alternatively, the
invariance can be recovered by considering a mirror image in the hidden
sector. In contrast, CP violation is much more profound because it
implies the violation of time reversal (T) invariance under the
assumption of TCP invariance. 
When this T violation is transmitted to the gravity sector, parity violation would
occur in the gravity sector because it is legitimate to require the C and
TCP invariance of the gravity sector. In general relativity, this
transmission can occur only through higher-order processes and is
usually suppressed. Though, this transmission might be substantial  
in a yet to be found ultra-violet (UV) completion of general
relativity, or the unified theory may allow the parity violation itself in
the gravity sector.  
For example, such phenomena indeed occur in superstring theories. Thus, experimental or
observational detections of parity violation in the gravity sector
certainly 
provide us with direct information concerning the UV completion of general relativity
or the ultimate unified theory.

The most promising route to this end would be to explore parity
violation in relics of inflation in the early universe. In particular,
measurements of parity violation in the primordial gravitational waves
produced during inflation are expected to bring us information about the
Planck scale physics.  From this standpoint, it has been intensively discussed how to
observe the parity violation in the power spectrum directly using the
laser interferometer~\cite{Seto:2006hf,Seto:2007tn,Seto:2008sr} and
indirectly using $TB$ correlation in the cosmic microwave background 
(CMB)~\cite{Saito:2007kt,Gluscevic:2010vv}. 
Remarkably, some speculative gravitational theories with a parity violating
 term  produce significant parity violation in the power
spectrum of primordial gravitational
waves~\cite{Contaldi:2008yz,Takahashi:2009wc}.  Unfortunately, in the
conventional inflationary scenario a parity violating term leads to only
very small amount of  circular polarization in the power spectrum of primordial gravitational
waves~\cite{Lue:1998mq,Choi:1999zy,Alexander:2004wk,Alexander:2006mt,Lyth:2005jf,Satoh:2007gn,Satoh:2008ck,Satoh:2010ep,Sorbo:2011rz}.

In principle, we can also seek parity violation in  higher-order
correlation functions~\cite{Maldacena:2002vr,Kamionkowski:2010rb}. Since
the power spectrum and the higher-order statistics
are often sensitive to different kinds of parity violating interactions,
they may have different correlations with other revealing
statistical features.  For example,  in a recent paper of Maldacena and
Pimentel, they discussed graviton non-gaussianity in the exact de Sitter spacetime and
found  that the pattern of parity violating bispectrum is severely constrained by
the conformal invariance~\cite{Maldacena:2011nz}.  One might then be tempted
to deduce that their result  implies the unsuppressed parity-violation in the
bispectrum. We argue that this is not immediately obvious since
they have not calculated observable quantities explicitly.

In this paper, we develop a new tool to analyze graviton correlation
functions and show that parity violation does {\it not} show up in
the case of the exact de Sitter spacetime. We will make clear the origin
of the apparent discrepancy between our result and that  by Maldacena
and Pimentel~\cite{Maldacena:2011nz}. 
Furthermore, by extending the analysis to the slow-roll
inflation, we find that parity violation in the bispectrum does not
vanish and its magnitude is proportional to the slow roll parameter.  
Although  the parity violation we found is small in the case of  slow-roll
inflationary scenario,  it can  become significant in the non-standard
inflations such as 
Dirac-Born-Infeld inflation~\cite{Silverstein:2003hf,Alishahiha:2004eh}.  
 Much larger parity
violation is also expected to occur in  
the recently developed effective field theory 
approaches~\cite{Creminelli:2006xe,Cheung:2007st,Weinberg:2008hq} to which
our formulation can be also applicable.

The organization of the present paper is as follows. 
In section~\ref{sec:formalism}, we present a
useful formalism for evaluating non-gaussianity in the helicity
basis. In section~\ref{sec:dS} we calculate graviton bispectrum in the exact de
Sitter universe and explicitly show that no parity violation arises in
the non-gaussianity. In section~\ref{sec:slowroll}, we explain how to calculate graviton
bispectrum in the case of slow-roll inflation. In particular, we demonstrate that
parity violation in the non-gaussianity exists and is proportional to
the slow roll parameter. We also discuss observability of parity
violation in the conventional inflation and more general theories.  The
final section is devoted to conclusion. 
In Appendix~\ref{app:A}, we give details of our formalism. In Appendix~\ref{app:pol}, we
present useful formulas for polarization tensors.

%-----------------------------------------------%
%                                               %
%                Formulation                    %
%                                               %
%-----------------------------------------------%

%T1>Formulation
\section{A Formalism for Graviton Non-gaussianity}
\label{sec:formalism}

In this section, with the aid of  the helicity basis,  we present a
useful method to evaluate graviton non-gaussianity generated by a
parity-violating Weyl cubic term.  

Let us start with the Friedmann-Lema\^itle-Robertson-Walker (FLRW) metric
\Eq{
  ds^2 = a^2 (\eta) \left[ -d\eta^2 + \delta_{ij} dx^i dx^j \right] \ ,
}
where $i,j$ are indices of the spatial coordinates. 
No distinction is made between their upper and lower indices hereafter
for the three-dimensional tensorial quantities. 
Tensor perturbations on this background universe are defined by
\Eq{
  ds^2 = a^2 (\eta) \left[ -d\eta^2 
            + \left( \delta_{ij} + h_{ij} \right) dx^i dx^j \right] \ ,
\label{h_FLRW}
}
where $h_{ij}$ obeys the transverse traceless conditions
$h_{ii}=h_{ij,j}=0$.  
The gravitational action for the tensor perturbation reads
\Eq{
  S_G = \frac{1}{4\kappa^2} \int d\eta d^3x a^2 \left[ 
             \frac{1}{2} h'_{ij} h'_{ij} - \frac{1}{2} h_{ij,k}h_{ij,k} \right]\,,
}
where the prime denotes the differentiation with respect to the conformal
time $\eta $ and 
$\kappa^2 = 8\pi G$ with the Newton constant $G$.  
We have two physical degrees of freedom for tensor perturbations which
can be characterized by the symmetric polarization tensors
$e^{(\pm)}_{ij}(\bm{k})$ satisfying 
\Eq{
e^{(s)}_{ii} (\bm{k}) = 0 \ , \qquad k_j  e^{(s)}_{ij} (\bm{k}) = 0 \ ,
}
where $\bm{k}$ is a comoving wavenumber vector, 
and $s=\pm $ represents  the helicity states $\pm 2$.  Namely, they satisfy
\Eq{
     \epsilon_{ijl} \frac{\partial}{\partial x_l}   
     \left[ e^{(s)}_{mj} (\bm{k}) e^{i\bm{k}\cdot \bm{x}} \right]
     = s k e^{(s)}_{im} (\bm{k}) e^{i\bm{k}\cdot \bm{x}} \,,
}
with $k=|\bm{k}|$.  It is convenient to adopt the normalization
\Eq{
  e^{(s)}_{ij} (\bm{k})  e^{*(s')}_{ij} (\bm{k}) = \delta_{ss'} \ ,
\label{basis_prop1}
}
where $*$ represents a complex conjugate.  
Remark that if we choose the phase of each polarization tensor appropriately,  the following relations hold:
\Eq{
e^{*(s)}_{ij}(\bm{k})=e^{(-s)}_{ij}(\bm{k})=e^{(s)}_{ij}(-\bm{k}).
\label{basis_prop2}
}
These relations will play a crucial r\^ole in proving that no parity violation occurs in non-gaussianity for the exact de Sitter universe.

Now, it is straightforward to quantize tensor perturbations.
The mode expansion is written as
\Eq{
  h_{ij} (\bm{x},\eta) = 2\kappa \int \frac{d^3 k}{(2\pi)^{3/2} \sqrt{2k}}
  \sum_{s=\pm} \left[ e^{(s)}_{ij} (\bm{k}) u_k (\eta)  a_s (\bm{k}) 
             + e^{*(s)}_{ij} (-\bm{k}) u_k (\eta)^* a^\dagger_s (-\bm{k})  
              \right] e^{i\bm{k}\cdot \bm{x}}  \ ,
}
where the creation and annihilation operators are normalized as 
\Eq{
  \left[ a_s (\bm{k}) , a^\dagger_s (\bm{k}') \right] 
  =\delta_{ss'} \delta (\bm{k}-\bm{k}' ) \ .
\label{aad}
}
The mode function $u_k$ satisfies the evolution equation
\Eq{
  u''_k + 2 \frac{a'}{a} u'_k + k^2 u_k = 0 \ .
}
The canonical commutation relation leads to the
normalization condition
\Eq{
  u^*_k \frac{\partial}{\partial \eta} u_k 
  -  u_k \frac{\partial}{\partial \eta} u^*_k  = -\frac{2ik}{a^2} \ .
\label{modefunction:normalization}}
Once a set of  mode functions satisfying this normalization is
specified,  the corresponding Fock vacuum is determined  by $a_s(\bm{k})|0\rangle=0$. Then,  we can calculate graviton
correlation functions for each interaction with the help of the standard perturbation
technique. 

The purpose of this work is to explore the possibility of parity
violation. It is well known that the information of gravitational waves
is completely encoded in the Weyl tensor $W_{\mu\nu\lambda\rho}$. When
we restrict ourselves to pure gravity, possible parity-violating
interaction terms can be found easily.  Apparently, the linear term in
the Weyl tensor 
\Eq{
  \int d\eta d^3 x \sqrt{-g} \epsilon^{\mu\nu\lambda\rho}
                           W_{\mu\nu\lambda\rho}
}
vanishes, whereas the quadratic term
\Eq{
  \int d\eta d^3 x \sqrt{-g} \epsilon^{\mu\nu\lambda\rho}
                       W_{\mu\nu}{}^{\alpha\beta}W_{\alpha\beta\lambda\rho}
}
is a topological term.\footnote{
This is related to a Pontryagin number 
due to 
$R_{\mu\nu\lambda\rho}*R^{\mu\nu\lambda\rho}=W_{\mu\nu\lambda\rho}*W^{\mu\nu\lambda\rho}$ 
\cite{Grumiller:2007rv}.
} 
Thus, the first two terms in
$W_{\mu\nu\lambda\rho }$ are irrelevant to the
parity violation. It follows that 
the leading term comes from the following cubic action
\Eq{
   S_{\rm PV} = - b \int d\eta d^3 x \sqrt{-g} \epsilon^{\mu\nu\lambda\rho}
    W_{\mu\nu}{}^{\alpha\beta} W_{\alpha\beta}{}^{\gamma\delta} 
    W_{\lambda\rho\gamma\delta} \,,
\label{PVaction}
}
which is P- and T-odd. 
Here, $b$ is a constant with dimension ${\rm [length]}^2$
(in  $c=\hbar =1$ units).
We shall evaluate the graviton non-gaussianity generated by this term
and examine if the parity violation emerges in the bispectrum.

We first establish an efficient formulation to calculate graviton correlation
functions in a general FLRW universe. In this method,  the fact that the helicity decomposition is
related to the decomposition by a pseudo-duality in the Minkowski
spacetime plays an important r\^ole.  To see this, let us consider
gravitational waves $\gamma_{{\rm (M)}ij}$ in the Minkowski spacetime
(see Appendix~\ref{app:A}).  
In terms of the new
variable defined by 
\Eq{
 \gamma^{\pm}_{{\rm (M)}ij} = \frac{1}{2} \left( \gamma'_{{\rm (M)}ij}  
                         \mp i  \epsilon_{jkl}\gamma_{{\rm (M)}ik,l} \right) \ ,
}
which is transverse and symmetric, 
we can express the cube of the chiral combinations \eqref{Wpm:def} of
the Weyl tensor as $(W^\pm)^3 = 64 (\gamma_{\rm (M)}^\pm )^3$.  This remarkable
relation makes computations quite simple.  

When we consider gravitational waves $h_{ij}$ in an expanding universe, however, 
this simple result does not hold anymore.   Nevertheless, as is shown in the Appendix A,
if we  define $\gamma_{ij}$ in terms of the tensor perturbation $h_{ij}$
in an expanding universe~(\ref{h_FLRW}) as 
\Eq{
\gamma'_{ij} : = a h'_{ij} \ ,
 \label{miracle}
}
we obtain a useful relation
\Eq{
   W^\mu{}_{\nu\lambda\rho} (h) 
   = \frac{1}{a} W^\mu{}_{\nu\lambda\rho} (\gamma)\big|_{\rm Minkowski} \ ,
}
where the Weyl tensor on the right-hand side is understood to take the
same functional form as that for tensor perturbations in the Minkowski
spacetime. Since the algebraic structure of the Weyl tensor is the same
as that in the Minkowski background, if we define $\gamma^\pm_{ij}$ by 
\Eq{
 \gamma^{\pm}_{ij} := \frac{1}{2} \left( \gamma'_{ij}  
                         \mp i  \epsilon_{jkl}\gamma_{ik,l} \right) \ ,
}
the relation $(W^\pm)^3 \propto (\gamma^\pm )^3$ continues to hold
in the FLRW universe.  
Therefore,  we can write  the parity-violating cubic action~(\ref{PVaction}) 
in a general FLRW background in terms of these new perturbation variables $\gamma^{\pm}_{ij}$ as
\Eq{
  S_{\rm PV} = 8ib \int d\eta d^3 x a^{-5}
  \left[ (\gamma^{ +}_{ij})'(\gamma^{+}_{jk})'(\gamma^{+}_{ki})'
  -  (\gamma^{-}_{ij})'(\gamma^{ -}_{jk})' (\gamma^{-}_{ki})' \right]
   \ .
   \label{main-action}
}
From this, we can read off the interaction Hamiltonian
$H_{\rm PV}(\eta )$ as
\begin{align}
  H_{\rm PV} =& -\frac{8ib}{(2\pi )^6} a^{-5}  \int d^3 k_1 d^3 k_2 d^3 k_3 
  \delta (\bm{k}_1 + \bm{k}_2 + \bm{k}_3 ) \nonumber\\
 & \times \left[ \left( \gamma^{ +}_{ij}(\eta,\bm{k}_1 ) \right)'
  \left( \gamma^{ +}_{jk} (\eta,\bm{k}_2 ) \right)'
  \left( \gamma^{ +}_{ki} (\eta,\bm{k}_3 ) \right)'
  - \left( \gamma^{ -}_{ij} (\eta,\bm{k}_1 ) \right)'
  \left(\gamma^{ -}_{jk} (\eta,\bm{k}_2 ) \right)'
  \left(\gamma^{ -}_{ki}(\eta,\bm{k}_3 ) \right)' \right]
   \,.
&
\label{main-hamiltonian}
\end{align}
We should emphasize that 
$\gamma_{ij}=\gamma^{+}_{ij}+\gamma^{-}_{ij}$ does not represent
 gravitational waves in Minkowski spacetime but an auxiliary field.
 From the definition (\ref{miracle}), we can give an explicit mode
 expansion expression for   $\gamma^\pm_{ij}$ as: 
\Eqr{
  \left( \gamma^{\pm}_{ij}\right)' 
  &=& \kappa \int \frac{d^3 k}{(2\pi)^{3/2} \sqrt{2k}}
  \sum_{s=\pm} \left[ e^{(s)}_{ij} (\bm{k}) 
  \left( \partial_\eta  \mp isk \right)\left( a u'_k (\eta) \right) a_s (\bm{k}) 
  \right. \nonumber\\
 &&  \left.   {\hskip 4cm}
  + e^{*(s)}_{ij} (-\bm{k})  \left( \partial_\eta  \mp isk \right)
             \left( a u'_k (\eta) \right)^* a^\dagger_s (-\bm{k})  
              \right] e^{i\bm{k}\cdot \bm{x}}  \ ,
 \label{main-relation}
}
Practically, this expression can be regarded as the definition of
$\gamma^{\pm}_{ij}$. Note that it is generally impossible to give a
closed expression for $\gamma^{\pm}_{ij}$. Nevertheless, it does not cause
any problem because only its time derivative
$\inpare{\gamma^{\pm}_{ij}}'$ appears in the interaction Hamiltonian~(\ref{main-hamiltonian}). 
 
Remarkably, in the exact de Sitter case, we see that $\gamma_{ij}$ has a
formal similarity to that in Minkowski spacetime  and the decomposition
$\gamma^\pm_{ij}$ corresponds to the helicity decomposition. Note that, in 
general cases,  $\gamma^+_{ij}$ and $\gamma^-_{ij}$ include 
both helicity components. The
main utility of our formalism is that a concise expression (\ref{main-action})
considerably simplifies calculations of bispectrum in a general expanding universe.

%-----------------------------------------------%
%                                               %
%                  de Sitter                    %
%                                               %
%-----------------------------------------------%

%T1>de Sitter case
\section{No Parity Violating Bispectrum in de Sitter Universe}
\label{sec:dS}

Employing the formulation developed in the previous section
let us compute the bispectrum in de Sitter universe.  We  
prove that no parity violation appears in non-gaussianity 
due to the symmetry of de Sitter universe. 

The mode function in a de Sitter background reads
\Eq{
  u_k = \frac{H}{k} \left( 1 + ik\eta \right) e^{-ik\eta } \ ,
}
where $H$ is a constant Hubble parameter. 
Letting
\begin{align}
\gamma _{ij}^\pm (\eta , \bm x)
= \int \frac{d ^3 k}{(2\pi)^3}\gamma^\pm_{ij} (\eta, \bm k) e^{i \bm k \cdot
 \bm x} \,, 
\end{align}
and substituting the above mode function into~(\ref{main-relation}), 
we obtain the following formula
\Eq{
   \left( \gamma^{+}_{ij} (\eta ,\bm{k}) \right)'
   = 2i \kappa k (2\pi)^{3/2} \sqrt{\frac{k}{2}}
   \left[ e^{(+)}_{ij}(\bm{k}) e^{-ik\eta } a_+ (\bm{k}) 
  - e^{*(-)}_{ij}(-\bm{k})  e^{ik\eta } a^\dagger_- (-\bm{k}) \right]
   \,,
\label{gammap_ds}
}
and
\Eq{
   \left( \gamma^{-}_{ij} (\eta ,\bm{k}) \right)'
   = 2i \kappa k (2\pi)^{3/2} \sqrt{\frac{k}{2}}
   \left[ e^{(-)}_{ij}(\bm{k}) e^{-ik\eta } a_- (\bm{k}) 
  - e^{*(+)}_{ij}(-\bm{k})  e^{ik\eta } a^\dagger_+ (-\bm{k}) \right] \,.
\label{gammam_ds}
}
Here, the operator $\gamma^{+}_{ij}$ annihilates a helicity $+2$ graviton
and creates a helicity $-2$ graviton. While, the operator $\gamma^{-}_{ij}$
does the opposite. 
In the de Sitter background it is easy to integrate these equations. 
In the asymptotic limit $\eta =0$, we have
\Eqrsubl{gamma0}{
 \gamma^{+}_{ij} (0,\bm{p}) 
     &=& -2\kappa (2\pi)^{3/2} \sqrt{\frac{p}{2}}
   \left[ e^{(+)}_{ij}(\bm{p}) a_+ (\bm{p}) 
     + e^{*(-)}_{ij}(-\bm{p})   a^\dagger_- (-\bm{p}) \right] \,,
\\
\gamma^{-}_{ij} (0,\bm{p})
    & =& -2\kappa (2\pi)^{3/2} \sqrt{\frac{p}{2}}
   \left[ e^{(-)}_{ij}(\bm{p}) a_- (\bm{p}) 
     + e^{*(+)}_{ij}(-\bm{p})   a^\dagger_+ (-\bm{p}) \right] \,.
}
Thus, we arrive at the simple relation~\footnote{Since $h_{ij}$ and
  $\gamma_{ij}$ are related via the differential
  equation~(\ref{miracle}),  $\gamma^{\pm}_{ij}$ at
  $\eta=0$ can take arbitrary value. 
  Hence,  Eq.~\eqref{gamma0} should be regarded simply as the definition of $\gamma^{\pm}_{ij}(0,\bm{p})$.}
\Eq{
 h_{ij} (0,\bm{p} ) = - \frac{H}{p^2}  \left[
        \gamma^+_{ij} (0,\bm{p}) + \gamma^-_{ij} (0,\bm{p}) \right] \ .
\label{h0bygamma}}

It is worth noting here the following property
\begin{align}
 \gamma_{ij}^\pm (\eta , \bm k)= \left(\gamma_{ij}^\mp (\eta , -\bm
 k)\right)^\dagger \,.
\end{align}
This traces back to the property of helicity basis~(\ref{basis_prop2}) 
and plays a key r\^ole in proving no parity violation in the exactly de
Sitter universe.

In the calculations of bispectrum based on 
the in-in formalism, we need
\begin{align}
  \EXP{ 0|\gamma^{\pm}_{ij} (0,\bm{p})  \left( \gamma^\mp_{kl} (\eta
 ,\bm{k}) \right)' |0} 
  &= 2 i \kappa^2 (2\pi)^3 k^2 \Pi^{\pm}_{ij,kl} (\bm{p})
              \delta (\bm{k} + \bm{p}) e^{ik\eta}  \,, \\
  \EXP{0|\left( \gamma^\mp_{kl} (\eta ,\bm{k}) \right)' \gamma^\pm_{ij} (0,\bm{p})  |0 }
  &= - 2i \kappa^2 (2\pi)^3 k^2 \Pi^{\mp}_{ij,kl} (\bm{p})
              \delta (\bm{k} + \bm{p}) e^{-ik\eta}  \,, \\
  \EXP{0|\gamma_{ij}^\pm (0, \bm p) (\gamma_{kl}^\pm (\eta , \bm k))'
 |0}
& = \EXP{0|(\gamma_{kl}^\pm (\eta , \bm k))'\gamma_{ij}^\pm (0, \bm p) 
 |0}=0 \,,
\end{align}
where we have introduced the projection operators
\Eq{
  \Pi^{\pm}_{ij,kl} (\bm{p}) 
  =  e^{(\pm)}_{ij}(\bm{p}) e^{*(\pm)}_{kl}(\bm{p}) \ .
}
Given these ingredients, we can compute the bispectrum as
\Eqr{
&& \EXP{  \gamma^{+}_{i_1 j_1} (0,\bm{p}_1)  
       \gamma^{+}_{i_2 j_2} (0,\bm{p}_2) 
       \gamma^{+}_{i_3 j_3} (0,\bm{p}_3)    } \nonumber\\
 &&  = i \int^0_{-\infty} d\eta
        \EXP{0|  \left[ H_{\rm PV}(\eta ) , \ \gamma^{+}_{i_1 j_1} (0,\bm{p}_1)  
       \gamma^{+}_{i_2 j_2} (0,\bm{p}_2) 
       \gamma^{+}_{i_3 j_3} (0,\bm{p}_3) \right]      |0} \nonumber\\  
 && =  384  ib \kappa^6 H^5 (2\pi)^3  p_1^2 p_2^2 p_3^2
      \delta (\bm{p}_1 + \bm{p}_2 + \bm{p}_3 )  
      \frac{5!}{(p_1 + p_2 + p_3 )^6}\nonumber\\
 && \quad   \times \left[
      \Pi^+_{i_1 j_1 ,kl} (\bm{p}_1)  
      \Pi^+_{i_2 j_2 ,lm} (\bm{p}_2)  \Pi^+_{i_3 j_3 ,mk} (\bm{p}_3)
     + \Pi^-_{i_1 j_1 ,kl} (\bm{p}_1)  
      \Pi^-_{i_2 j_2 ,lm} (\bm{p}_2)  \Pi^-_{i_3 j_3 ,mk} (\bm{p}_3)
       \right] \,.
      \label{plus}
}
where we have used the formula
\Eq{
\int^0_{-\infty} d\eta \eta^{5} 
 e^{i(p_1 + p_2 + p_3 )\eta}  
=  \frac{5!}{(p_1 + p_2 + p_3 )^6}
      \ .
}
The other non-zero contribution is given by
\Eqr{
&& \EXP{  \gamma^{-}_{i_1 j_1} (0,\bm{p}_1)  
       \gamma^{-}_{i_2 j_2} (0,\bm{p}_2) 
       \gamma^{-}_{i_3 j_3} (0,\bm{p}_3)    } \nonumber\\
 &&  = i\int^0_{-\infty} d\eta
  \EXP{0|  \left[  H_{\rm PV} (\eta) , \ \gamma^{-}_{i_1 j_1} (0,\bm{p}_1)  
       \gamma^{-}_{i_2 j_2} (0,\bm{p}_2) 
       \gamma^{-}_{i_3 j_3} (0,\bm{p}_3) \right]      |0} \nonumber\\  
 && = -  384 i b \kappa^6 H^5 (2\pi)^3  p_1^2 p_2^2 p_3^2
      \delta (\bm{p}_1 + \bm{p}_2 + \bm{p}_3 ) 
      \frac{5!}{(p_1 + p_2 + p_3 )^6} \nonumber\\
 && \quad   \times \left[
      \Pi^-_{i_1 j_1 ,kl} (\bm{p}_1)  
      \Pi^-_{i_2 j_2 ,lm} (\bm{p}_2)  \Pi^-_{i_3 j_3 ,mk} (\bm{p}_3)
     + \Pi^+_{i_1 j_1 ,kl} (\bm{p}_1)  
      \Pi^+_{i_2 j_2 ,lm} (\bm{p}_2)  \Pi^+_{i_3 j_3 ,mk} (\bm{p}_3)
      \right]\,.
      \label{minus}
}
It can be also verified that 
the mixed parts $\langle \gamma^+\gamma^+\gamma^-\rangle$
and $\langle \gamma^+\gamma^+\gamma^-\rangle$ vanish, as they should due
to the symmetry of de Sitter spacetime.

These results are in agreement with those obtained by Maldacena and Pimentel~\cite{Maldacena:2011nz}.
They argued that this is determined by the conformal symmetry of de Sitter spacetime.
It is worthwhile to emphasize that neither $\langle(\gamma^+)^3\rangle$ nor
$\langle(\gamma^-)^3\rangle$  themselves are direct observables.   

Now, 
we need to clarify if the parity violation is observable or not in the non-gaussianity.
Let us define the right-handed and left-handed circular polarizations by
\begin{align}
 &h^R := h_{ij} e^{*(+)}_{ij}  \,, \qquad 
h^L := h_{ij} e^{*(-)}_{ij} \,,
\end{align}
respectively. A possible observable quantity in which a parity violation is
encoded is their difference
$\langle(h^R)^3\rangle-\langle(h^L)^3\rangle$. 
Here we have 
\begin{align}
& \EXP{h^{R,L} (0,\bm{p}_1 ) h^{R,L} (0,\bm{p}_2 ) h^{R,L} (0,\bm{p}_3
 )} 
\nonumber \\
  &= -\frac{H^3}{p_1^2 p_2^2 p_3^2}
 e^{*(\pm)}_{i_1 j_1}(\bm{p}_1 ) 
 e^{*(\pm)}_{i_2 j_2} (\bm{p}_2 ) e^{*(\pm)}_{i_3 j_3} (\bm{p}_3 )
\nonumber\\ & \quad \times
\langle(\gamma_{i_1 j_1}^+(0,\bm{p}_1)+\gamma_{i_1 j_1}^-(0,\bm{p}_1))
(\gamma_{i_2 j_2}^+(0,\bm{p}_2) +\gamma_{i_2 j_2}^-(0,\bm{p}_2) )(
       \gamma_{i_3 j_3}^+(0,\bm{p}_3)+ \gamma_{i_3 j_3}^- (0,\bm{p}_3))
 \rangle 
\nonumber \\
  &= -\frac{H^3}{p_1^2 p_2^2 p_3^2}
 e^{*(\pm)}_{i_1 j_1}(\bm{p}_1 ) 
 e^{*(\pm)}_{i_2 j_2} (\bm{p}_2 ) e^{*(\pm)}_{i_3 j_3} (\bm{p}_3 )
\nonumber\\ & \quad 
\times \left(
\EXP{\gamma_{i_1 j_1}^+ (0,\bm{p}_1)  \gamma_{i_2 j_2}^+ (0,\bm{p}_2) 
       \gamma_{i_3 j_3}^+ (0,\bm{p}_3) } 
+\EXP{\gamma_{i_1 j_1}^- (0,\bm{p}_1)  \gamma_{i_2 j_2}^- (0,\bm{p}_2) 
       \gamma_{i_3 j_3}^- (0,\bm{p}_3) }
\right)   \,.
\end{align}
For explicit formula, it is convenient to define a function~(see
Appendix~\ref{app:pol} for derivation of the last equality)
\Eqr{
 F(p_1 ,\  p_2 , \ p_3 ) 
 &:=&  e^{*(+)}_{kl} (\bm{p}_1 ) e^{*(+)}_{lm} (\bm{p}_2 ) 
 e^{*(+)}_{mk} (\bm{p}_3 ) \nonumber\\
 & =&  e^{*(-)}_{kl} (\bm{p}_1 ) e^{*(-)}_{lm} (\bm{p}_2 ) 
 e^{*(-)}_{mk} (\bm{p}_3 ) \nonumber\\
 &=& - \frac{\left( p_1 + p_2 + p_3 \right)^3 
  \left(p_1 +p_2 -p_3 \right)\left( p_2 + p_3 - p_1 \right)
  \left(p_3 + p_1 -p_2  \right)}{64 p_1^2 p_2^2 p_3^2} \ . 
  \label{polarization}
}
Then, we obtain the bispectrum of right handed circular polarized modes
\Eqr{
&& \EXP{h^{R} (0,\bm{p}_1 ) h^{R} (0,\bm{p}_2 ) h^{R} (0,\bm{p}_3 )}
\nonumber \\
&& \quad =  
  - 384 ib (2\pi)^3  \kappa^6   H^8 
       F(p_1 ,\  p_2 , \ p_3 )  \left[
    \frac{5!}{(p_1 + p_2 + p_3 )^6}  
    -  \frac{5!}{(p_1 + p_2 + p_3 )^6} \right]
    \delta (\bm{p}_1 + \bm{p}_2 + \bm{p}_3 )  \nonumber \\
&& \quad =0 \,,
}
and the bispectrum of left-handed circular polarized modes
\Eqr{
&& \EXP{h^{L} (0,\bm{p}_1 ) h^{L} (0,\bm{p}_2 ) h^{L} (0,\bm{p}_3 )}
\nonumber \\
&&  \quad=  
  384 ib (2\pi)^3  \kappa^6   H^8 
       F(p_1 ,\  p_2 , \ p_3 )  \left[
    \frac{5!}{(p_1 + p_2 + p_3 )^6} 
    - \frac{5!}{(p_1 + p_2 + p_3 )^6} \right]
    \delta (\bm{p}_1 + \bm{p}_2 + \bm{p}_3 ) \nonumber \\ 
&&        \quad                     = 0  \ .
}
In the above calculations, the contributions from (\ref{plus}) and (\ref{minus})
have been cancelled out.
It turns out that  there exists no parity violation
in a pure de Sitter universe. 
Since Maldacena and Pimentel proved that $*W W^2 $ 
is the only way to break parity in the bispectrum,
we have established that no parity violating bispectrum
exists in exact de Sitter spacetime.

%-----------------------------------------------%
%                                               %
%                  slow roll                    %
%                                               %
%-----------------------------------------------%

%T1>Slow roll inflation
\section{Parity Violation during Slow Roll Inflation}
\label{sec:slowroll}

In this section, we extend calculations in the previous section 
to a slow-rolling inflationary universe.
The strategy is as follows.
We already know there is no parity violation in the pure de Sitter case.
Hence, when we expand every term up to the first order in the slow roll parameter,
the sources of parity violation stem from the following three parts:

\begin{itemize}
\item[(i)] change in the asymptotic mode function
\item[(ii)] change of $\gamma^\pm_{ij}$
\item[(iii)] change of the cosmic expansion
\end{itemize}
In what follows we shall discuss these contributions separately. 
We finally put each contribution together to conclude that parity violation
appears in the first-order in slow roll.  
We will see that the contribution (i) is higher-order in slow-roll, i.e.,
it is insignificant.

%T2>mode function
\subsection{Change in the asymptotic mode function}

In the slow-roll stage, the universe undergoes the following evolution
\Eq{
ds^2 = -dt^2 + a^2 (t) \delta_{ij} dx^i dx^j
     = a^2 (\eta) \left[ -d\eta^2 + \delta_{ij} dx^i dx^j  \right] \,.
}
Though the explicit functional form of the scale factor is sensitive to the
inflaton potential and/or kinetic term, 
the scale factor takes a simple form in the leading order of the slow-roll
parameter.  To see this, let us define the slow roll parameter
$\epsilon$ as\footnote{This is slightly abuse of nomenclature since the
inflaton may not be slowly rolling the potential. The parameter
$\epsilon $ simply measures the departure from constancy of $H$. 
}
\Eq{
\epsilon = - \frac{\dot{H}}{H^2} = - \frac{H'}{aH^2} \ , \qquad
H= \frac{\dot{a}}{a}  = \frac{a'}{a^2}  \,,
}
where the dot denotes the derivative with respect to the cosmic time $t$.
In the leading order of the slow-roll approximation this slow-roll
parameter $\epsilon $ can be regarded as a constant.
Under this condition by integrating  
\Eq{
  \left( \frac{1}{aH} \right)' = \epsilon -1,
}
which is equivalent to the definition of  $\epsilon$, 
we obtain 
\Eq{
 a (\eta ) = \left( -H_* \eta \right)^{-1/(1-\epsilon)}
           = \left( -H_* \eta \right)^{1/2 -\nu} \ ,
}
where $H_*$ is a constant of integration, and we have defined
\Eq{
\nu = \frac{3}{2} + \frac{\epsilon}{1-\epsilon} \simeq \frac{3}{2} +\epsilon \ .
}
The equation of motion for gravitational waves in this background 
can be written as
\Eq{
h''_{ij} - \frac{2}{(1-\epsilon)\eta } h'_{ij} + k^2 h_{ij} =0 \ .
}
By solving this, we find that the mode function which satisfies the normalization condition~\eqref{modefunction:normalization} and approaches the Bunch-Davis type mode in the $\eta\tend-\infty$ limit  is given by 
\Eq{
  u_k (\eta) = \sqrt{\frac{\pi k}{2H_*}} e^{i\pi \nu /2 - i\pi /4}
  \left( -H_* \eta \right)^\nu H_\nu^{(1)} \left( -k\eta \right) \ ,
}
where $H_\nu^{(1)} $ is the Hankel function of the first kind.
In the asymptotic limit $-\eta \rightarrow 0$, this function freezes
out and approaches to the constant value
\Eq{
  u_k (\eta ) = \frac{H_*}{k} e^{i\pi \epsilon /2}
  \left[ 1+ \epsilon \left( 2-\gamma -\log 2 + \log\frac{H_*}{k} \right) \right] \ ,
}  
where $\gamma$ is the Euler constant. Here, we used 
\Eq{
H_\nu^{(1)} (z) \sim -  \frac{i}{\pi} \Gamma (\nu) \left( \frac{z}{2} \right)^{-\nu}
\ ,\quad {\rm as}\  |z|\ll 1  \quad {\rm for \quad Re}[\nu] >0 \ .
}
and $\Gamma' (z) = \psi (z) \Gamma(z)$, where $\psi (z) $ is  poly-Gamma
function.  
If we absorb the phase into the definition of the ladder operators as 
 $a_{(s)} (\bm{k}) \rightarrow  e^{-i\pi \epsilon /2} a_{(s)} (\bm{k})$,
we find that a relation between $h_{ij}$ and $\gamma_{ij}$ at
$\eta=0$  is the same as that in de Sitter case except for a real-valued numerical factor: 
\Eq{
 h_{ij} (0,\bm{p} ) = - \frac{H_*}{p^2} C \left[
        \gamma^+_{ij} (0,\bm{p}) + \gamma^-_{ij} (0,\bm{p}) \right] \ ,
}
where 
\Eq{
C=1+ \epsilon \left( 2-\gamma -\log 2 + \log H_* /k \right)\,.
\label{C:def}
}
As it turns out, this term does not contribute to the parity-violation
in the first-order in slow-roll parameter.

%T2>Int. Hamiltonian
\subsection{Change of $\gamma^\pm_{ij}$}

Next, let us consider the change in the expression for the interaction
Hamiltonian in terms of the creation-annihilation operators. In the
lowest order in the slow roll parameter $\epsilon$, this change is
caused by the change of $\gamma^\pm_{ij}$ from that in
the pure de Sitter case in the linear order in $\epsilon$.   In this
section, we denote $\gamma^\pm_{ij}$ in the slow roll inflation
background by $\b\gamma^\pm_{ij}$  to distinguish it from the
corresponding quantity in the pure de Sitter background, which is
denoted by $\gamma^\pm_{ij}$ without the bar. 
 
Now we have
\begin{align}
  \left[ \partial_\eta -isk \right] \left( a u_k' (\eta) \right)
  =& k^2 e^{i\pi \nu /2 - i\pi /4} \sqrt{\frac{\pi }{2}}
  \left[ \left( \frac{3}{2}-\nu \right) \left( -k\eta \right)^{-1/2}
   H_{\nu-1}^{(1)} \left( -k\eta \right) \right. \nonumber\\
 &  {\hskip 2cm} \left. 
 + \left( -k\eta \right)^{1/2} H_{\nu-2}^{(1)} \left( -k\eta \right)
   +i s \left( -k\eta \right)^{1/2} H_{\nu-1}^{(1)} \left( -k\eta \right)
   \right] \ .
\end{align}
Inserting $\nu \simeq 3/2 +\epsilon$, we can approximate it as
\Eqr{
 && \left[ \partial_\eta -isk \right] \left( a u_k' (\eta) \right) \nonumber\\
 && \quad =  ik^2 e^{i\pi \epsilon /2} \left[
     (1+s) e^{-ik\eta}   
     + \epsilon 
  \left\{ \frac{i}{-k\eta} e^{-ik\eta}
    - \frac{i\pi}{2} (1+s) e^{-ik\eta } 
     -  (1-s) e^{ik\eta} E_1 (2ik\eta ) \right\}
  \right]  \ , \notag\\
&&
}
where 
\Eq{
E_1 (z) = \int^\infty_z \frac{e^{-t}}{t} dt  \ ,
}
and we have used the formulas
\Eqrsub{
\left.
 \frac{\partial H_{\nu}^{(1)}(z) }{\partial \nu} \right|_{\nu=1/2}
 &=& \sqrt{\frac{2}{\pi z}} \left[  -i E_1 (-2iz) e^{-iz} - \frac{\pi}{2} e^{iz} \right]\,,
\\
 \left.\frac{\partial H_{\nu}^{(1)}(z) }{\partial \nu} \right|_{\nu=-1/2}
 &=& \sqrt{\frac{2}{\pi z}} \left[ -E_1 (-2iz) e^{-iz} 
 -i \frac{\pi}{2} e^{iz}  \right]
 \ . 
}
Absorbing the phase factor $ e^{i\pi \epsilon /2}$ by  rescaling,
we have
\Eqr{
\left( \bar{\gamma}_{ij}^+ \right)'
&=& \kappa \int \frac{d^3 k}{(2\pi)^{3/2} \sqrt{2k}}
\left[ e^{(+)}_{ij}(\bm{k})  \left[ \partial_\eta -ik \right]
 \left( a u_k' (\eta) \right) a_+ (\bm{k}) \right. \nonumber\\
 &&    {\hskip 3cm}  \left. 
  + e^{(+)*}_{ij}(-\bm{k})  \left[ \partial_\eta -ik \right]
 \left( a u_k' (\eta) \right)^{*} a^\dagger_+ (-\bm{k})
 \right. \nonumber\\
 &&    {\hskip 3cm}  \left. 
  + e^{(-)}_{ij}(\bm{k})  \left[ \partial_\eta + ik \right]
 \left( a u_k' (\eta) \right) a_- (\bm{k})
 \right. \nonumber\\
 &&    {\hskip 3cm}  \left. 
  + e^{(-)*}_{ij}(-\bm{k})  \left[ \partial_\eta +ik \right]
 \left( a u_k' (\eta) \right)^{*} a^\dagger_- (-\bm{k})
\right] e^{i\bm{k}\cdot \bm{x}} \\
&=& \left( \gamma_{ij}^+ \right)'
 + \epsilon \chi^{+}_{ij} \ , \nonumber 
}
where $\left( \gamma_{ij}^+ \right)' $ is the expression  in the pure de
Sitter case given in (\ref{gammap_ds}) and (\ref{gammam_ds}).  
In terms of the functions $\rho_\pm$ defined by
\Eqrsub{
\rho_{+} (k,\eta) &=&  i \frac{1}{-k\eta} e^{-ik\eta}
  -i \pi e^{-ik\eta}  \ , \\
\rho_{-} (k,\eta) &=&  i \frac{1}{-k\eta} e^{-ik\eta}
       - 2 {E_1} (2ik\eta ) e^{ik\eta}
    \,,
}
the correction term $\chi^\pm_{ij}$ can be  expressed as
\begin{align}
\chi^{\pm}_{ij} =&
  ik^2 \kappa \int \frac{d^3 k}{(2\pi)^{3/2} \sqrt{2k}}
\left[ e^{(+)}_{ij}(\bm{k})  \rho_\pm (k,\eta)
 a_+ (\bm{k}) 
  - e^{(+)*}_{ij}(-\bm{k})  \rho_{\mp} (k,\eta)^* a^\dagger_+ (-\bm{k})
 \right. \nonumber\\
 &    {\hskip 3.5cm}  \left.  
  + e^{(-)}_{ij}(\bm{k})  \rho_{\mp} (k,\eta) a_- (\bm{k})
  - e^{(-)*}_{ij}(-\bm{k})  \rho_{\pm} (k,\eta)^* a^\dagger_- (-\bm{k})
\right] e^{i\bm{k}\cdot \bm{x}} \ .
\end{align}

The tensor $\chi^\pm_{ij}$ describing the deviation of
$\bar{\gamma}_{ij}$  away from that in the pure de Sitter case appears in the interaction Hamiltonian as
\Eqr{
   H_{\rm PV} &=& - ib \epsilon \int d^3 x  a^{-5}
  24 \left[  \left( \gamma^{+}_{ij} \right)' \left( \gamma^{+}_{jk} \right)'
               \chi^{+}_{ki}  -  {\rm h.c}. \right]  \nonumber\\
    &=&  -  ib \epsilon \frac{24}{(2\pi)^6} \int d^3 k_1 d^3 k_2 d^3 k_3 
   \delta (\bm{k}_1 + \bm{k}_2 + \bm{k}_3 ) \nonumber\\
&& \qquad \times    a^{-5}
   \left[  \left( \gamma^{+}_{ij} (\eta , \bm{k}_1)\right)' 
   \left( \gamma^{+}_{jk} (\eta , \bm{k}_2)\right)' 
               \chi^{+}_{ki} (\eta , \bm{k}_3)  \right. \nonumber\\
        && \left.  {\hskip 3.5cm}
             -  \left( \gamma^{-}_{ij} (\eta , \bm{k}_1)\right)' 
   \left( \gamma^{-}_{jk} (\eta , \bm{k}_2)\right)' 
               \chi^{-}_{ki} (\eta , \bm{k}_3)  \right]            \,,
} 
where h.c. denotes the hermite conjugation of the preceding term.

Now, we can calculate the bispectrum in the slow-roll inflation due to the change of $\gamma^{\pm}_{ij}$.  From the structure of $\chi^\pm_{ij}$, we see that graviton non-gaussianity  consists of  the four parts: $\EXP{\gamma^+ \gamma^+ \gamma^+ }$ ,  $ \EXP{\gamma^- \gamma^- \gamma^- } $,  $\EXP{\gamma^+ \gamma^+ \gamma^- } $,
and  $\EXP{\gamma^- \gamma^- \gamma^+ } $.   We calculate these parts in
order using the Wightman functions, 
\Eqrsub{
 && \EXP{0|  \gamma^\pm_{ij} (0,\bm{p}) \chi^\mp_{kl} (\eta ,\bm{k})  |0}
  = i \kappa^2 (2\pi)^3 k^2 \Pi^{\pm}_{ij,kl} (\bm{p})
              \delta (\bm{k} + \bm{p}) \rho^{*}_{+} (p,\eta ) \,,\\
&&  \EXP{0| \chi^\mp_{kl} (\eta ,\bm{k}) \gamma^\pm_{ij} (0,\bm{p})   |0}
  = - i \kappa^2 (2\pi)^3 k^2 \Pi^{\mp}_{ij,kl} (\bm{p})
              \delta (\bm{k} + \bm{p}) \rho_{+} (p,\eta )  \,,
}
and 
\Eqrsub{
&&   \EXP{0|  \gamma^\pm_{ij} (0,\bm{p}) \chi^\pm_{kl} (\eta ,\bm{k})  |0}
  = i \kappa^2 (2\pi)^3 k^2 \Pi^{\pm}_{ij,kl} (\bm{p})
              \delta (\bm{k} + \bm{p}) \rho^{*}_{-} (p,\eta ) \,,
\\
&&  \EXP{0|  \chi^\pm_{kl} (\eta ,\bm{k})  \gamma^\pm_{ij} (0,\bm{p}) |0}
  = -i \kappa^2 (2\pi)^3 k^2 \Pi^{\mp}_{ij,kl} (\bm{p})
              \delta (\bm{k} + \bm{p}) \rho_{-} (p,\eta )  \ .
}
%

%T3>+++
\subsubsection{$(+++)$ part}

This part is evaluated as 
\begin{align}
& \EXP{  \gamma^{+}_{i_1 j_1} (0,\bm{p}_1)  
       \gamma^{+}_{i_2 j_2} (0,\bm{p}_2) 
       \gamma^{+}_{i_3 j_3} (0,\bm{p}_3)    } \nonumber\\
 &  = i \int^0_{-\infty} d\eta 
   \EXP{0|  \left[  H_{\rm PV} (\eta ) , \ \gamma^{+}_{i_1 j_1} (0,\bm{p}_1)  
       \gamma^{+}_{i_2 j_2} (0,\bm{p}_2) 
       \gamma^{+}_{i_3 j_3} (0,\bm{p}_3) \right]      |0} \nonumber\\  
 & = 96 ib \epsilon  \kappa^6 H_*^5 (2\pi)^3  p_1^2 p_2^2 p_3^2
      \delta (\bm{p}_1 + \bm{p}_2 + \bm{p}_3 )  \nonumber\\
 & \quad   \times \left[
      \Pi^+_{i_1 j_1 ,kl} (\bm{p}_1)  
      \Pi^+_{i_2 j_2 ,lm} (\bm{p}_2)  \Pi^+_{i_3 j_3 ,mk} (\bm{p}_3)
     A_+ 
     + \Pi^-_{i_1 j_1 ,kl} (\bm{p}_1)  
      \Pi^-_{i_2 j_2 ,lm} (\bm{p}_2)  \Pi^-_{i_3 j_3 ,mk} (\bm{p}_3)
      A_+^* \right]   \nonumber \\
& \quad +\textrm{5 permutations}
      \,,  
      \label{gamma-plus}
\end{align}
where $A_+$ is the integral
\Eqr{
A_+ &=& \int^0_{-\infty} d\eta \eta^{5} 
 e^{i(p_1 + p_2 )\eta}  \rho^*_+ (p_3 ,\eta ) \nonumber \\ 
 &=& \int^0_{-\infty} d\eta \eta^5  e^{i(p_1 + p_2 )\eta}
 \left[ -i \frac{1}{-p_3 \eta}e^{ip_3 \eta}
           +i\pi e^{ip_3 \eta}  \right] \nonumber\\
&=& \frac{ 4!}{p_3 ( p_1 +p_2 + p_3 )^5} 
      + i\pi \frac{5!}{(p_1 + p_2 + p_3 )^6}
      \ .
}
In contrast to the exact de Sitter case, this integral gives rise to an imaginary
 part.
This produces a crucial difference in the final result regarding the parity
 violation.  Here, we should stress that both of the circular
polarization modes contribute to the $(+++)$-part of bispectrum.  

%T3>---
\subsubsection{$(---)$ part}

This part can be obtained from  $\EXP{\gamma^+ \gamma^+ \gamma^+ }$ by
complex conjugation and the flip of the directions of each momentum $\bm{p}_i$: 
\begin{align}
& \EXP{  \gamma^{-}_{i_1 j_1} (0,\bm{p}_1)  
       \gamma^{-}_{i_2 j_2} (0,\bm{p}_2) 
       \gamma^{-}_{i_3 j_3} (0,\bm{p}_3)    } \nonumber\\
 &  = i \int^0_{-\infty} d\eta
  \EXP{0|  \left[  H_{\rm PV} (\eta) , \ \gamma^{-}_{i_1 j_1} (0,\bm{p}_1)  
       \gamma^{-}_{i_2 j_2} (0,\bm{p}_2) 
       \gamma^{-}_{i_3 j_3} (0,\bm{p}_3) \right]      |0} \nonumber\\  
 & = -96 ib \epsilon  \kappa^6 H_*^5 (2\pi)^3  p_1^2 p_2^2 p_3^2
      \delta (\bm{p}_1 + \bm{p}_2 + \bm{p}_3 )  \nonumber\\
 & \quad   \times \left[
      \Pi^-_{i_1 j_1 ,kl} (\bm{p}_1)  
      \Pi^-_{i_2 j_2 ,lm} (\bm{p}_2)  \Pi^-_{i_3 j_3 ,mk} (\bm{p}_3)
     A_+ 
     + \Pi^+_{i_1 j_1 ,kl} (\bm{p}_1)  
      \Pi^+_{i_2 j_2 ,lm} (\bm{p}_2)  \Pi^+_{i_3 j_3 ,mk} (\bm{p}_3)
      A_+^* \right]  \nonumber \\
& \quad +\textrm{5 permutations}
      \,.
      \label{gamma-minus}
\end{align}
Here, again, we see that both polarization modes make nonvanishing contributions to the bispectrum.

%T3>--+
\subsubsection{$(--+)$ part}

In contrast to the pure de Sitter case, we have the cross contributions
\Eqr{
&& \EXP{  \gamma^{-}_{i_1 j_1} (0,\bm{p}_1)  
       \gamma^{-}_{i_2 j_2} (0,\bm{p}_2) 
       \gamma^{+}_{i_3 j_3} (0,\bm{p}_3)    } \nonumber\\
 &&  = i \int^0_{-\infty} d\eta 
   \EXP{0|  \left[  H_{\rm PV} (\eta) , \ \gamma^{-}_{i_1 j_1} (0,\bm{p}_1)  
       \gamma^{-}_{i_2 j_2} (0,\bm{p}_2) 
       \gamma^{+}_{i_3 j_3} (0,\bm{p}_3) \right]      |0} \nonumber\\  
 && = -96 ib \epsilon  \kappa^6 H_*^5 (2\pi)^3  p_1^2 p_2^2 p_3^2
      \delta (\bm{p}_1 + \bm{p}_2 +  \bm{p}_3 )  \nonumber\\
 && \quad   \times \left[
      \Pi^-_{i_1 j_1 ,kl} (\bm{p}_1)  
      \Pi^-_{i_2 j_2 ,lm} (\bm{p}_2)  \Pi^+_{i_3 j_3 ,mk} (\bm{p}_3)
     A_- 
     + \Pi^+_{i_1 j_1 ,kl} (\bm{p}_1)  
      \Pi^+_{i_2 j_2 ,lm} (\bm{p}_2)  \Pi^-_{i_3 j_3 ,mk} (\bm{p}_3)
      A_-^* \right]  \nonumber\\
  && \qquad + \left\{ (i_1 j_1 ) \leftrightarrow (i_2 j_2 ) \right\}
      \ . 
}
Here, $A_-$  is the integral 
\Eqr{
A_- &=& \int^0_{-\infty} d\eta \eta^{5} 
 e^{i(p_1 + p_2 )\eta}  \rho^*_- (p_3 ,\eta ) \nonumber\\
 &=& \int^0_{-\infty} d\eta \eta^5  e^{i(p_1 + p_2 )\eta}
 \left[ -i \frac{1}{-p_3 \eta}e^{ip_3 \eta}
           -2 E_1 (-2ip_3\eta) e^{-ip_3 \eta}  \right] \nonumber\\
&=& \frac{ 4!}{p_3 ( p_1 +p_2 + p_3 )} 
      -2 K_5
      \ ,
}
where
\Eq{
K_5 = \int_{-\infty}^0 d\eta \eta^5 e^{i(p_1 + p_2 - p_3 ) \eta }
         \int_{-2ip_3 \eta}^\infty \frac{e^{-t}}{t}  dt  \ .
}
In this expression, let us set $p_1 +p_2 - p_3=E$ and regard $E$ as a
constant independent of $p_3$. Differentiation of the corresponding
expression gives  
\Eqr{
\frac{\partial}{\partial p_3} K_5 
&=& -\frac{1}{p_3}\int_{-\infty}^0 d\eta \eta^5 e^{i(2 p_3 + E) \eta }  \nonumber\\
&=& -\frac{1}{p_3} \frac{5!}{(E + 2p_3)^6}  \nonumber\\
&=&  \frac{\partial^5}{\partial E^5}  \frac{1}{p_3} \frac{1}{(E + 2p_3)} \nonumber\\
&=&   \frac{\partial^5}{\partial E^5}  \frac{\partial}{\partial p_3}
        \frac{1}{E}\log \left( \frac{2{p_3}}{E+2p_3}\right) \ .
}
Since the integral must vanish in the limit $p_3 \rightarrow i \infty$,  
this equation can be integrated to yield
\Eq{
  K_5 =   \frac{\partial^5}{\partial E^5}  \inrbra{
        \frac{1}{E}\log \left( \frac{2{p_3}}{E+2p_3}\right) } \ .
}
Thus, it turned out that $A_-$ contains no imaginary part in contrast to $A_+$.

%T3>++-
\subsubsection{$(++-)$ part}

In parallel with the previous one, 
this part of  the bispectrum reads
\Eqr{
&& \EXP{  \gamma^{+}_{i_1 j_1} (0,\bm{p}_1)  
       \gamma^{+}_{i_2 j_2} (0,\bm{p}_2) 
       \gamma^{-}_{i_3 j_3} (0,\bm{p}_3)    } \nonumber\\
 &&  = i \int^0_{-\infty} d\eta
 \EXP{0|  \left[  H_{\rm PV} (\eta) , \ \gamma^{+}_{i_1 j_1} (0,\bm{p}_1)  
       \gamma^{+}_{i_2 j_2} (0,\bm{p}_2) 
       \gamma^{-}_{i_3 j_3} (0,\bm{p}_3) \right]      |0} \nonumber\\  
 && = 96 ib \epsilon  \kappa^6 H_*^5 (2\pi)^3  p_1^2 p_2^2 p_3^2
      \delta (\bm{p}_1 + \bm{p}_2 +  \bm{p}_3 )  \nonumber\\
 && \quad   \times \left[
      \Pi^+_{i_1 j_1 ,kl} (\bm{p}_1)  
      \Pi^+_{i_2 j_2 ,lm} (\bm{p}_2)  \Pi^-_{i_3 j_3 ,mk} (\bm{p}_3)
     A_- 
     + \Pi^-_{i_1 j_1 ,kl} (\bm{p}_1)  
      \Pi^-_{i_2 j_2 ,lm} (\bm{p}_2)  \Pi^+_{i_3 j_3 ,mk} (\bm{p}_3)
      A_-^* \right] \nonumber\\
  && \qquad + \left\{ (i_1 j_1 ) \leftrightarrow (i_2 j_2 ) \right\}
      \ .  
}
%

%T2>cosmic expansion
\subsection{Change of the cosmic expansion}

The final contribution comes from the imaginary part of the integral
\begin{align}
  \int^0_{-\infty} d\eta (-\eta)^{5+5\epsilon} e^{-i(p_1+p_2+p_3)\eta} 
&=\frac{\Gamma(6+5\epsilon)}{(-i)^{6+5\epsilon} (p_1+p_2+p_3)^{6+5\epsilon}}
\nonumber \\
&  \sim  -\frac{5!}{(p_1+p_2+p_3)^6} \left( 1+ \frac{5\pi}{2} i\epsilon   \right) 
\equiv - \frac{5!}{(p_1+p_2+p_3)^6} - \epsilon B \,, 
\end{align}
where the real part of $O(\epsilon )$ has been neglected in the second
line since it has
nothing to do with the parity violation.
In terms of this integral, $\EXP{\gamma^+\gamma^+\gamma^+}$
 and $\EXP{\gamma^-\gamma^-\gamma^-}$ are expressed as
\begin{align}
& \EXP{  \gamma^{+}_{i_1 j_1} (0,\bm{p}_1)  
       \gamma^{+}_{i_2 j_2} (0,\bm{p}_2) 
       \gamma^{+}_{i_3 j_3} (0,\bm{p}_3)    } \nonumber\\
 & = {384} ib \epsilon  \kappa^6 H_*^5 (2\pi)^3  p_1^2 p_2^2 p_3^2
      \delta (\bm{p}_1 + \bm{p}_2 +  \bm{p}_3 )  \nonumber\\
 & \quad   \times \left[
      \Pi^+_{i_1 j_1 ,kl} (\bm{p}_1)  
      \Pi^+_{i_2 j_2 ,lm} (\bm{p}_2)  \Pi^+_{i_3 j_3 ,mk} (\bm{p}_3)
     B ^*
     + \Pi^-_{i_1 j_1 ,kl} (\bm{p}_1)  
      \Pi^-_{i_2 j_2 ,lm} (\bm{p}_2)  \Pi^-_{i_3 j_3 ,mk} (\bm{p}_3)
      B \right] \,,
      \label{expansion-plus}
\end{align}
and
\begin{align}
& \EXP{  \gamma^{-}_{i_1 j_1} (0,\bm{p}_1)  
       \gamma^{-}_{i_2 j_2} (0,\bm{p}_2) 
       \gamma^{-}_{i_3 j_3} (0,\bm{p}_3)    } \nonumber\\
 & = - 384 ib \epsilon  \kappa^6 H_*^5 (2\pi)^3  p_1^2 p_2^2 p_3^2
      \delta (\bm{p}_1 + \bm{p}_2 +  \bm{p}_3 )  \nonumber\\
 & \quad   \times \left[
      \Pi^-_{i_1 j_1 ,kl} (\bm{p}_1)  
      \Pi^-_{i_2 j_2 ,lm} (\bm{p}_2)  \Pi^-_{i_3 j_3 ,mk} (\bm{p}_3)
     B^* 
     + \Pi^+_{i_1 j_1 ,kl} (\bm{p}_1)  
      \Pi^+_{i_2 j_2 ,lm} (\bm{p}_2)  \Pi^+_{i_3 j_3 ,mk} (\bm{p}_3)
      B \right] 
      \,.  
      \label{expansion-minus}
\end{align}
From these expressions, it is found that the imaginary part of $B$ fails to 
cancel the contributions to the non-gaussianity from 
$\EXP{\gamma^+\gamma^+\gamma^+}$ and $\EXP{\gamma^-\gamma^-\gamma^-}$.

%-----------------------------------------------%
%                                               %
%              Parity violation                 %
%                                               %
%-----------------------------------------------%

%T2>parity violation
\subsection{Parity Violation}

Now, we are in a position to discuss parity violation in a slow-roll
inflationary universe by assembling results obtained above. 
To see the parity violation, we are required to evaluate 
\begin{align}
& \EXP{h^{R,L} (0,\bm{p}_1 ) h^{R,L} (0,\bm{p}_2 ) h^{R,L} (0,\bm{p}_3 )} \nonumber\\
&\quad 
=-\frac{H_*^3 C^3}{p_1^2 p_2^2 p_3^2}
 e^{*(\pm)}_{i_1 j_1}(\bm{p}_1 ) 
 e^{*(\pm)}_{i_2 j_2} (\bm{p}_2 ) e^{*(\pm)}_{i_3 j_3} (\bm{p}_3 )
\nonumber \\
& \qquad \times
\left(
\EXP{\gamma_{i_1 j_1}^+ (0,\bm{p}_1)  \gamma_{i_2 j_2}^+ (0,\bm{p}_2) 
\gamma_{i_3 j_3}^+ (0,\bm{p}_3)}+\EXP{\gamma_{i_1 j_1}^+ (0,\bm{p}_1)\gamma_{i_2 j_2}^- (0,\bm{p}_2) 
\gamma_{i_3 j_3}^- (0,\bm{p}_3)}
\right)
\,.
\end{align}
In the previous subsections, we found that the three-point correlators
of $\gamma^\pm_{ij} $ are of order $\epsilon $. It follows that the change from the overall constant factor $C$  
becomes higher-order in $\epsilon $.  Thus we set $C=1$ henceforth.
From Eqs.~(\ref{gamma-plus}), (\ref{gamma-minus}), (\ref{expansion-plus}),
and (\ref{expansion-minus}), we obtain the final result
\Eqr{
&& \EXP{h^{R} (0,\bm{p}_1 ) h^{R} (0,\bm{p}_2 ) h^{R} (0,\bm{p}_3 )} \nonumber\\
&& \qquad =  
  - 32 ib \epsilon  (2\pi)^3   \kappa^6   H_*^8 
       F(p_1 ,\  p_2 , \ p_3 )  \left[
   3  A_+  -  3  A_+^* -2 B +2 B^* \right]
   \delta (\bm{p}_1 + \bm{p}_2 + \bm{p}_3 ) \nonumber\\
   && \qquad\qquad    + {\rm 5 \ permutations}  \nonumber\\
&& \qquad = -64 (2\pi)^4 \epsilon b \kappa^6   H_*^8 
       \delta (\bm{p}_1 + \bm{p}_2 + \bm{p}_3 ) F(p_1 ,\  p_2 , \ p_3 ) 
  \frac{6!}{(p_1 + p_2 + p_3 )^6} \,,
}
and 
\Eqr{
&& \EXP{h^{L} (0,\bm{p}_1 ) h^{L} (0,\bm{p}_2 ) h^{L} (0,\bm{p}_3 )} \nonumber\\
&&  \qquad=  
  32 ib \epsilon (2\pi)^3   \kappa^6   H_*^8 
       F(p_1 ,\  p_2 , \ p_3 )  \left[
    3  A_+  - 3  A_+^* -2 B +2 B^* \right]\delta (\bm{p}_1 + \bm{p}_2 + \bm{p}_3 ) 
   \nonumber\\
   && \qquad\qquad    + {\rm 5 \ permutations}  \nonumber\\
&& \qquad =  64 (2\pi)^4 \epsilon b \kappa^6   H_*^8  
       \delta (\bm{p}_1 + \bm{p}_2 + \bm{p}_3 ) F(p_1 ,\  p_2 , \ p_3 ) 
        \frac{6!}{(p_1 + p_2 + p_3 )^6} \ ,
}
where $F(p_1, p_2, p_3)$ has been defined in Eq.~(\ref{polarization}).
Apparently, parity violation shows up in the bispectrum and
its magnitude is proportional to the slow roll parameter.
It is interesting to observe that the bispectrum of
curvature perturbations is also proportional to the slow-roll
parameter in the conventional single inflationary scenario~\cite{Maldacena:2002vr}.

It should be noted that no parity violation occurs in the mixed parts.
Introducing a function
\begin{align}
J(p_1,p_2,p_3):= e^{*(+)}_{kl} (\bm{k}_1 ) e^{*(+)}_{lm} (\bm{k}_2 ) 
 e^{*(-)}_{mk} (\bm{k}_3 ) 
  =  e^{*(-)}_{kl} (\bm{k}_1 ) e^{*(-)}_{lm} (\bm{k}_2 ) 
 e^{*(+)}_{mk} (\bm{k}_3 )  \,,
\label{polorizationJ}
\end{align}
a simple algebra shows that 
\Eqr{
&& \EXP{h^{L} (0,\bm{p}_1 ) h^{L} (0,\bm{p}_2 ) h^{R} (0,\bm{p}_3 )} \nonumber\\
&&  \qquad=  
  i 192 (2\pi)^3  \epsilon b \kappa^6   H_*^8 
       J(p_1 ,\  p_2 , \ p_3 )  \left[
    A_-  - A_-^* \right]\delta (\bm{p}_1 + \bm{p}_2 + \bm{p}_3 )\,. 
}
Since $A_-$ is real, there exists no parity violating contribution.
The same conclusion applies to the bispectrum $\EXP{h^R h^R h^L}$.

It should be stressed that the parity violation can be observed in the CMB.
In fact, three-point correlators $\EXP{TTB}, \EXP{TEB}, \EXP{EEB}$ become non-zero in contrast to the parity 
conserving cases. In the conventional slow-roll inflationary scenario, the amplitude
might be too small to be detected in near future. However, in the non-conventional
scenarios, we might have much larger parity violation.

%T1>Conclusion
\section{Conclusion}
\label{sec:conclusion}

We have developed a useful formalism to evaluate graviton correlation
functions. As an illustrating application of this formalism, we studied parity violation in the early
universe through non-gaussianity of gravitons. First of all, by
calculating the bispectrum we have shown that  no parity violation
arises in the exact de Sitter background. This statement may appear to
be inconsistent with that by Maldacena and Pimentel.  However, the
explicit results of calculations  they gave in their paper are in
agreement with ours. The difference came from the interpretation. They
have calculated $\EXP{ (\gamma^{\pm} )^3 } $ and concluded there is a
parity violation. This is not the case since the actual observables are 
$\EXP{(h^R)^3}-\EXP{(h^L)^3 }$, etc. 
These vanish in the purely de Sitter universe.  
The situation is different when the spacetime departs from the exact de
Sitter. 
In slow-roll inflationary case we have found parity
violation in the graviton bispectrum proportional to the slow roll
parameter.  

We also discussed that parity violation in the bispectrum can be
observed e.g., in the $\langle TTB\rangle$ correlation in the CMB.  It might be also possible
to detect the signature of parity violation through direct observations
of primordial gravitational waves using a space interferometer
observatory~\cite{Taruya:2006kqa}. 

Although we have concentrated on the parity violating Weyl cubic term,
the present formalism enables us to calculate higher-order parity violating correlation functions
rather straightforwardly. In fact, it does not require a 
sophisticated  expertise to evaluate the
parity violating non-gaussianity of the type $*W W^n$. We can also
extend our analysis to more general inflationary scenario using the
effective field theory approach. In those cases, we expect large parity
violating non-gaussianity.  Moreover, we can evaluate parity violating
contribution due to a non-minimal coupling to the inflation $\phi$ such
as a term $f(\phi) *W W$~\cite{Weinberg:2008hq}. Our formalism would be useful in analyzing
these general cases.  Work along these directions is in progress.

It is known that exact de Sitter correlation function is related to
correlation function of stress tensor in the conformal field theory
through analytic continuation~\cite{Maldacena:2002vr}.  Hence, the graviton non-gaussianity has
an interesting application. Similarly, our results would be useful for
calculating three-point functions of stress tensor in the non-conformal
field theory using gravity/field theory correspondence.

%T1>Ack
\section*{Acknowledgements}
We would like to thank Takahiro Tanaka and Yoshihisa Kitazawa for
discussions. We are grateful to Xian Gao for useful comments. 
This work is supported by the Grant-in-Aid for  Scientific Research Fund of the Ministry of  Education, Science and Culture of Japan No.22540274, the Grant-in-Aid for Scientific Research (A) (No.21244033, No.22244030), the Grant-in-Aid for  Scientific Research on Innovative Area No.21111006, JSPS under the Japan-Russia Research Cooperative Program and the Grant-in-Aid for the Global COE Program ``The Next Generation of Physics, Spun from Universality and Emergence.''

%T1>appendix
\appendix

%T2>chiral decomposition formulae
\section{Pseudo Self-duality, Helicity,  Parity Violating Action}
\label{app:A}

The pseudo self-dual and anti-self-dual Weyl tensor are defined by
\Eq{
W^{\pm}_{\mu\nu\lambda\sigma}:= W_{\mu\nu\lambda\sigma}\pm i * W_{\mu\nu\lambda\sigma},\quad
 * W^{\mu\nu}{}_{\lambda\rho} 
     = \frac{1}{2} \epsilon^{\mu\nu\alpha\beta} W_{\alpha \beta \lambda\rho} \ ,
\label{Wpm:def}
}
which satisfy
\Eq{
   * W^{\pm}_{\mu\nu\lambda\rho} = \mp i W^{\pm}_{\mu\nu\lambda\rho} \ .
}
The factor ``$i$'' arises due to the Lorentzian signature. 
Since the relation 
$W^+_{\mu\nu}{}^{\alpha\beta} W^-_{\alpha\beta}{}^{\gamma\delta} =0$
holds, we get
\Eq{
\epsilon^{\mu\nu\lambda\rho}
    W_{\mu\nu}{}^{\alpha\beta} W_{\alpha\beta}{}^{\gamma\delta} 
    W_{\lambda\rho\gamma\delta}
 = \frac{1}{4i} \left(  W^+_{\mu\nu}{}^{\alpha\beta} 
                         W^+_{\alpha\beta}{}^{\gamma\delta}
                          W^+_{\gamma\delta}{}^{\mu\nu}
            - W^-_{\mu\nu}{}^{\alpha\beta} 
                   W^-_{\alpha\beta}{}^{\gamma\delta}  
                    W^-_{\gamma\delta}{}^{\mu\nu} \right)\ .
}
Note that we also have a relation
\Eq{
 W_{\mu\nu}{}^{\alpha\beta} W_{\alpha\beta}{}^{\gamma\delta} 
    W_{\gamma\delta\mu\nu}
 = \frac{1}{8} \left(  W^+_{\mu\nu}{}^{\alpha\beta} 
                         W^+_{\alpha\beta}{}^{\gamma\delta}
                          W^+_{\gamma\delta}{}^{\mu\nu}
            + W^-_{\mu\nu}{}^{\alpha\beta} 
                   W^-_{\alpha\beta}{}^{\gamma\delta}  
                    W^-_{\gamma\delta}{}^{\mu\nu} \right)\ .
}

Let us consider tensor perturbations in the Minkowski spacetime
\Eq{
ds^2 =  -d\eta^2 
            + \left( \delta_{ij} + \gamma_{{\rm (M)}ij} \right) dx^i dx^j  \ ,
}
where $\gamma_{{\rm (M)}ij}$ is a transverse traceless symmetric tensor
describing a gravitational wave. 
The on-shell equations of motion are
\Eq{
  - {\gamma}''_{{\rm (M)}ij} + \nabla^2 \gamma_{{\rm (M)}ij} =0 \,,
\label{eom_flat}
}
where the prime denotes the differentiation with respect to $\eta$. 
It is convenient to define a new variable by 
$\tilde{\gamma}_{{\rm (M)}ij}:= \epsilon_{jkl}\gamma_{{\rm (M)}ik,l}$. 
By multiplying $\epsilon_{mij}$ to both side, we can prove this new tensor
is symmetric. Then, it is easy to see the transversality of 
$\tilde{\gamma}_{{\rm (M)}ij}$.
Thus $\tilde{\gamma}_{{\rm (M)} ij}$ is also the transverse traceless symmetric tensor. 
Using this new tensor, we can define helicity eigenstate by
\Eq{
  \gamma^{\pm}_{{\rm (M)} ij} := \frac{1}{2} \left( \gamma'_{{\rm (M)}ij}
                   \mp i \tilde{\gamma}_{{\rm (M)}ij}  \right)\,.
}
One can then verify the relation
\Eq{
  \epsilon_{ijl} \frac{\partial}{\partial x_l}  \gamma^{\pm}_{{\rm (M)}mj} 
     = \pm k \gamma^{\pm}_{{\rm (M)}im} \,.
}
A direct calculation yields
\Eqr{
W^0{}_{i0k} &=&  \frac{1}{2} {\gamma}''_{{\rm (M)}ik} \\
W^0{}_{jkl} &=& -\gamma_{{\rm (M)}j[k,l]}  = -\frac{1}{2} \epsilon_{klm} \tilde{\gamma}_{{\rm (M)}jm} \\
W^j{}_{klm} &=& - \delta_{l[j} {\gamma}''_{{\rm (M)}k]m} 
                   + \delta_{m[j} {\gamma}''_{{\rm (M)}k]l} \ ,
}
where we used on-shell equations for gravitational waves~(\ref{eom_flat}) in Minkowski spacetime. 
It is easy to derive the following relations~\footnote{In appearance the expression of
$W^3$ derived here disagrees with Eq.~(2.14) in~\cite{Maldacena:2011nz}. 
However, they can be shown to be equivalent by using  
three-dimensional identity 
$2(\gamma'_{{\rm (M)}i[k,l]}\gamma'_{{\rm (M)}j[k,l]}+\gamma'_{{\rm (M)}k[l,i]}\gamma'_{{\rm (M)}k[l,j]})\equiv\delta_{ij}\gamma'_{{\rm (M)}kl,m}\gamma'_{{\rm (M)}k[l,m]}$. }

\Eq{
  W^\pm_{\mu\nu}{}^{\alpha\beta} W^\pm_{\alpha\beta}{}^{\gamma\delta} 
  {W^\pm_{\gamma\delta}}^{\mu\nu}
    = 64 (\gamma^\pm_{{\rm (M)}ij})' (\gamma^\pm_{{\rm (M)}jk})' (\gamma^\pm_{{\rm (M)}ki})' \ .
}
and 
\begin{align}
W_{\mu\nu}{}^{\alpha\beta} W_{\alpha\beta}{}^{\gamma\delta} 
  {W_{\gamma\delta}}^{\mu\nu}&=2\gamma_{{\rm (M)}ij}''
 \left[\gamma_{{\rm (M)}jk}''\gamma_{{\rm (M)}ki}''
-3\tilde{\gamma}_{{\rm (M)}jk}'\tilde\gamma_{{\rm (M)}ki}' \right]\,, \\
\epsilon^{\mu\nu\lambda\rho}
    W_{\mu\nu}{}^{\alpha\beta} W_{\alpha\beta}{}^{\gamma\delta} 
    W_{\lambda\rho\gamma\delta}
 &= 4\tilde\gamma_{{\rm (M)}ij}''
 \left[\tilde\gamma_{{\rm (M)}jk}''\tilde\gamma_{{\rm (M)}ki}''
-3\gamma_{{\rm (M)}jk}'\gamma_{{\rm (M)}ki}' \right]\,.
\end{align}
These equations manifest the symmetry between $W^3$ and $*WW^2$
in a flat space.

Now, we would like to promote the above beautiful relation to
FLRW spacetime~(\ref{h_FLRW}). 
The Weyl tensor is obtained as
\Eqr{
W^0{}_{i0k} &=&  \frac{1}{2a} \left(ah'_{ik} \right)' \\
W^0{}_{jkl} &=& - h'_{j[k,l]}
  \\
W^j{}_{klm} &=& \frac{1}{a}\left[ - \delta_{l[j} (ah'_{k]m})' 
                   + \delta_{m[j} (ah'_{k]l})' \right] \ ,
}
where  
we have used on-shell equations for gravitational waves in FLRW spacetime.
Comparing this expression with the one in Minkowski spacetime, we
notice that if we make the identification 
\Eq{
    \gamma'_{ij}  = a h'_{ij} \ ,
}
a simple relation
\Eq{
   W^\mu{}_{\nu\lambda\rho} (h) 
   = \frac{1}{a} W^\mu{}_{\nu\lambda\rho} (\gamma)\big|_{\rm Minkowski}
}
holds irrespective of the expansion history of the universe. 
Therefore, with a definition 
\Eq{
  \gamma^\pm_{ij}  = \frac{1}{2} \left( \gamma'_{ij}  
                         \mp i  \epsilon_{jkl}\gamma_{ik,l} \right) \ ,
}
we obtain
\Eq{
     W^\pm_{\mu\nu}{}^{\alpha\beta} (h) W^\pm_{\alpha\beta}{}^{\gamma\delta} (h)
    W^\pm_{\gamma\delta}{}^{\mu\nu} (h) 
    = \frac{64}{a^9} \gamma^{\prime\pm}_{ij}
    \gamma^{\prime\pm}_{jk} \gamma^{\prime\pm}_{ki} \ .
\label{Wpm_cube}
}
Thus, we can deduce the action in the main text.

%T2>polarization tensor
\section{Polarization tensors}
\label{app:pol}

We fix the representation of polarization tensors~\cite{McFadden:2011kk}.
Because of momentum conservation $\bm{k}_1 + \bm{k}_2 + \bm{k}_3 =0$, 
we can make all $\bm k_i$ lying on the ($x, y$)-plane without losing any generality.
It follows that  a triangle can be constructed as
\Eq{
\bm{k}_1 = k_1 \left( 1 , 0 , 0 \right) \ , \quad
\bm{k}_2 = k_2 \left( \cos \theta , \sin\theta , 0 \right) \ , \quad
\bm{k}_3 = k_3 \left( \cos \phi , \sin\phi , 0 \right) \ ,
}
where
\begin{align}
& \cos\theta = \frac{k_3^2 - k_1^2 -k_2^2 }{2k_1 k_2} \ ,\quad
  \sin\theta = \frac{\lambda}{2k_1 k_2} \ , \quad
  \cos\phi = \frac{k_2^2 - k_3^2 -k_1^2 }{2k_3 k_1} \ ,\quad
  \sin\phi = - \frac{\lambda}{2k_3 k_1} \ , \\
& \lambda = \sqrt{2k_1^2 k_2^2 + 2 k_2^2 k_3^2 + 2 k_3^2 k_1^2 
                    - k_1^4 -k_2^4 -k_3^4 }  \ .
\end{align}
Note that $0 \leq \theta \leq \pi$ and $\pi \leq \phi \leq 2\pi$.
Using the above representation, we can fix the polarization basis.
For $\bm{k}_1$, we have a simple expression
\Eq{
e^{(s_1)} (\bm{k}_1) 
= \frac{1}{2}\left( 
\begin{array}{ccc}
 0 & 0 & 0 \\
 0 & 1 & i s_1 \\
 0 & i s_1 & -1
\end{array}
\right) \ ,
}
where $s_1$ distinguishes the helicity.
To obtain the polarization for $\bm{k}_2$, we need to rotate the above one
by $\theta$ as
\Eq{
e^{(s_2)} (\bm{k}_2) 
= \frac{1}{2}\left( 
\begin{array}{ccc}
 \sin^2 \theta & -\sin \theta \cos\theta & -i s_2 \sin\theta \\
 -\sin\theta \cos\theta & \cos^2 \theta & i s_2 \cos\theta \\
 -i s_2 \sin \theta  & i s_2 \cos \theta & -1
\end{array}
\right) \ .
}
Similarly, we obtain the  polarization tensor for $\bm k_3$ as
\Eq{
e^{(s_3)} (\bm{k}_3) 
= \frac{1}{2}\left( 
\begin{array}{ccc}
 \sin^2 \phi & -\sin \phi \cos\phi & -i s_3 \sin\phi \\
 -\sin\phi \cos\phi & \cos^2 \phi & i s_3 \cos\phi \\
 -i s_3 \sin \phi  & i s_3 \cos \phi & -1
\end{array}
\right) \ .
}
From these expressions we can calculate 
$F(k_1 , k_2, k_3 )$ and $J(k_1 , k_2, k_3 )$ defined respectively by
(\ref{polarization}) and (\ref{polorizationJ}).
The resultant expressions are given by
\Eqr{
  F(k_1 , k_2 ,  k_3 ) &=& -\frac 18 
(1- \cos \theta ) (1-\cos \phi )(1- \cos(\phi-\theta))
  \nonumber\\
  &=& - \frac{\left\{ (k_1 +k_2 )^2 -k_3^2 \right\} 
  \left\{(k_2 +k_3 )^2 -k_1^2\right\} 
  \left\{ (k_3 +k_1 )^2 -k_2^2 \right\}}{64 k_1^2 k_2^2 k_3^2}   \nonumber\\
  &=& - \frac{\left( k_1 + k_2 + k_3 \right)^3 
  \left(k_1 +k_2 -k_3 \right)\left( k_2 + k_3 - k_1 \right)
  \left(k_3 + k_1 -k_2  \right)}{64 k_1^2 k_2^2 k_3^2}\,,
}
and 
\Eqr{
 J(k_1 ,  k_2 ,  k_3 ) 
&=& -\frac 18 (1-\cos \theta) (1+\cos\phi)
(1+\cos (\phi-\theta))  \nonumber \\
&  =& - \frac{\left( k_1 + k_2 + k_3 \right) 
  \left(k_1 +k_2 -k_3 \right)^3 \left( k_2 + k_3 - k_1 \right)
  \left(k_3 + k_1 -k_2  \right)}{64 k_1^2 k_2^2 k_3^2}   \ . 
\label{J:def}
}
%

%T1>References
%

%T1>End
\end{document}


\begin{thebibliography}{99}

%\cite{Seto:2006hf}
\bibitem{Seto:2006hf}
  N.~Seto,
  %``Prospects for direct detection of circular polarization of
  %gravitational-wave background,''
  Phys.\ Rev.\ Lett.\  {\bf 97}, 151101 (2006)
  [arXiv:astro-ph/0609504].
  %%CITATION = PRLTA,97,151101;%%
    
%\cite{Seto:2007tn}
\bibitem{Seto:2007tn}
  N.~Seto and A.~Taruya,
  %``Measuring a Parity Violation Signature in the Early Universe via
  %Ground-based Laser Interferometers,''
  Phys.\ Rev.\ Lett.\  {\bf 99}, 121101 (2007)
  [arXiv:0707.0535 [astro-ph]].
  %%CITATION = PRLTA,99,121101;%%
  
%\cite{Seto:2008sr}
\bibitem{Seto:2008sr}
  N.~Seto and A.~Taruya,
  %``Polarization analysis of gravitational-wave backgrounds from the
  %correlation signals of ground-based interferometers: measuring a
  %circular-polarization mode,''
  Phys.\ Rev.\  D {\bf 77}, 103001 (2008)
  [arXiv:0801.4185 [astro-ph]].
  %%CITATION = PHRVA,D77,103001;%%
  
%\cite{Saito:2007kt}
\bibitem{Saito:2007kt}
  S.~Saito, K.~Ichiki and A.~Taruya,
  %``Probing polarization states of primordial gravitational waves with CMB
  %anisotropies,''
  JCAP {\bf 0709}, 002 (2007)
  [arXiv:0705.3701 [astro-ph]].
  %%CITATION = JCAPA,0709,002;%%  
    
%\cite{Gluscevic:2010vv}
\bibitem{Gluscevic:2010vv}
  V.~Gluscevic and M.~Kamionkowski,
  %``Testing Parity-Violating Mechanisms with Cosmic Microwave Background
  %Experiments,''
  arXiv:1002.1308 [astro-ph.CO].
  %%CITATION = ARXIV:1002.1308;%%


%\cite{Contaldi:2008yz}
\bibitem{Contaldi:2008yz}
  C.~R.~Contaldi, J.~Magueijo and L.~Smolin,
  %``Anomalous CMB polarization and gravitational chirality,''
  Phys.\ Rev.\ Lett.\  {\bf 101}, 141101 (2008)
  [arXiv:0806.3082 [astro-ph]].
  %%CITATION = PRLTA,101,141101;%%
      
%\cite{Takahashi:2009wc}
\bibitem{Takahashi:2009wc}
  T.~Takahashi and J.~Soda,
  %``Chiral Primordial Gravitational Waves from a Lifshitz Point,''
  Phys.\ Rev.\ Lett.\  {\bf 102}, 231301 (2009)
  [arXiv:0904.0554 [hep-th]].
  %%CITATION = PRLTA,102,231301;%%
  
%\cite{Lue:1998mq}
\bibitem{Lue:1998mq}
  A.~Lue, L.~M.~Wang and M.~Kamionkowski,
  %``Cosmological signature of new parity-violating interactions,''
  Phys.\ Rev.\ Lett.\  {\bf 83}, 1506 (1999)
  [arXiv:astro-ph/9812088].
  %%CITATION = PRLTA,83,1506;%%
  
%\cite{Choi:1999zy}
\bibitem{Choi:1999zy}
  K.~Choi, J.~-c.~Hwang, K.~W.~Hwang,
  %``String theoretic axion coupling and the evolution of cosmic structures,''
  Phys.\ Rev.\  {\bf D61}, 084026 (2000).
  [hep-ph/9907244].
   
%\cite{Alexander:2004wk}
\bibitem{Alexander:2004wk}
  S.~Alexander and J.~Martin,
  %``Birefringent gravitational waves and the consistency check of  inflation,''
  Phys.\ Rev.\  D {\bf 71}, 063526 (2005)
  [arXiv:hep-th/0410230].
  %%CITATION = PHRVA,D71,063526;%%

%\cite{Alexander:2006mt}
\bibitem{Alexander:2006mt}
  S.~H.~S.~Alexander,
  %``Is cosmic parity violation responsible for the anomalies in the WMAP data?,''
  Phys.\ Lett.\  {\bf B660}, 444-448 (2008).
  [hep-th/0601034].


%\cite{Lyth:2005jf}
\bibitem{Lyth:2005jf}
  D.~H.~Lyth, C.~Quimbay and Y.~Rodriguez,
  %``Leptogenesis and tensor polarisation from a gravitational Chern-Simons
  %term,''
  JHEP {\bf 0503}, 016 (2005)
  [arXiv:hep-th/0501153].
  %%CITATION = JHEPA,0503,016;%%
  
%\cite{Satoh:2007gn}
\bibitem{Satoh:2007gn}
  M.~Satoh, S.~Kanno and J.~Soda,
  %``Circular Polarization of Primordial Gravitational Waves in String-inspired
  %Inflationary Cosmology,''
  Phys.\ Rev.\  D {\bf 77}, 023526 (2008)
  [arXiv:0706.3585 [astro-ph]].
  %%CITATION = PHRVA,D77,023526;%%
  
%\cite{Satoh:2008ck}
\bibitem{Satoh:2008ck}
  M.~Satoh and J.~Soda,
  %``Higher Curvature Corrections to Primordial Fluctuations in Slow-roll
  %Inflation,''
  JCAP {\bf 0809}, 019 (2008)
  [arXiv:0806.4594 [astro-ph]].
  %%CITATION = JCAPA,0809,019;%%
  
%\cite{Satoh:2010ep}
\bibitem{Satoh:2010ep}
  M.~Satoh,
  %``Slow-roll Inflation with the Gauss-Bonnet and Chern-Simons Corrections,''
  JCAP {\bf 1011}, 024 (2010)
  [arXiv:1008.2724 [astro-ph.CO]].
  %%CITATION = JCAPA,1011,024;%%

 %\cite{Sorbo:2011rz}
\bibitem{Sorbo:2011rz}
  L.~Sorbo,
  %``Parity violation in the Cosmic Microwave Background from a pseudoscalar inflaton,''
  JCAP {\bf 1106}, 003 (2011).
  [arXiv:1101.1525 [astro-ph.CO]].

%\cite{Maldacena:2002vr}
\bibitem{Maldacena:2002vr}
  J.~M.~Maldacena,
  %``Non-Gaussian features of primordial fluctuations in single field
  %inflationary models,''
  JHEP {\bf 0305}, 013 (2003)
  [arXiv:astro-ph/0210603].
  %%CITATION = JHEPA,0305,013;%%
    
%\cite{Kamionkowski:2010rb}
\bibitem{Kamionkowski:2010rb}
  M.~Kamionkowski and T.~Souradeep,
  %``The Odd-Parity CMB Bispectrum,''
  arXiv:1010.4304 [astro-ph.CO].
  %%CITATION = ARXIV:1010.4304;%%  
  
%\cite{Maldacena:2011nz}
\bibitem{Maldacena:2011nz}
  J.~M.~Maldacena and G.~L.~Pimentel,
  %``On graviton non-Gaussianities during inflation,''
  arXiv:1104.2846 [hep-th].
  %%CITATION = ARXIV:1104.2846;%%   

%\cite{Silverstein:2003hf}
\bibitem{Silverstein:2003hf}
  E.~Silverstein and D.~Tong,
  %``Scalar speed limits and cosmology: Acceleration from D-cceleration,''
  Phys.\ Rev.\  D {\bf 70}, 103505 (2004)
  [arXiv:hep-th/0310221].
  %%CITATION = PHRVA,D70,103505;%%
  
%\cite{Alishahiha:2004eh}
\bibitem{Alishahiha:2004eh}
  M.~Alishahiha, E.~Silverstein and D.~Tong,
  %``DBI in the sky,''
  Phys.\ Rev.\  D {\bf 70}, 123505 (2004)
  [arXiv:hep-th/0404084].
  %%CITATION = PHRVA,D70,123505;%%
      
%\cite{Creminelli:2006xe}
\bibitem{Creminelli:2006xe}
  P.~Creminelli, M.~A.~Luty, A.~Nicolis and L.~Senatore,
  %``Starting the Universe: Stable Violation of the Null Energy Condition and
  %Non-standard Cosmologies,''
  JHEP {\bf 0612}, 080 (2006)
  [arXiv:hep-th/0606090].
  %%CITATION = JHEPA,0612,080;%%
  
%\cite{Cheung:2007st}
\bibitem{Cheung:2007st}
  C.~Cheung, P.~Creminelli, A.~L.~Fitzpatrick, J.~Kaplan and L.~Senatore,
  %``The Effective Field Theory of Inflation,''
  JHEP {\bf 0803}, 014 (2008)
  [arXiv:0709.0293 [hep-th]].
  %%CITATION = JHEPA,0803,014;%%
  
%\cite{Weinberg:2008hq}
\bibitem{Weinberg:2008hq}
  S.~Weinberg,
  %``Effective Field Theory for Inflation,''
  Phys.\ Rev.\  D {\bf 77}, 123541 (2008)
  [arXiv:0804.4291 [hep-th]].
  %%CITATION = PHRVA,D77,123541;%%

\bibitem{Grumiller:2007rv}
  D.~Grumiller, N.~Yunes,
  %``How do Black Holes Spin in Chern-Simons Modified Gravity?,''
  Phys.\ Rev.\  {\bf D77}, 044015 (2008).
  [arXiv:0711.1868 [gr-qc]].


%\cite{Taruya:2006kqa}
\bibitem{Taruya:2006kqa}
  A.~Taruya,
  %``Probing anisotropies of gravitational-wave backgrounds with a space-based
  %interferometer III: Reconstruction of a high-frequency skymap,''
  Phys.\ Rev.\  D {\bf 74}, 104022 (2006)
  [arXiv:gr-qc/0607080].
  %%CITATION = PHRVA,D74,104022;%%

%\cite{McFadden:2011kk}
\bibitem{McFadden:2011kk}
  P.~McFadden and K.~Skenderis,
  %``Cosmological 3-point correlators from holography,''
  JCAP {\bf 1106}, 030 (2011)
  arXiv:1104.3894 [hep-th].
  %%CITATION = ARXIV:1104.3894;%%
    
       
\end{thebibliography}
\end{document}